\documentclass[%
 reprint,
superscriptaddress,
 amsmath,amssymb,
 aps,
prb,
]{revtex4-2}

\usepackage{graphicx}
\usepackage{dcolumn}
\usepackage{bm}
\usepackage[colorlinks,linkcolor=blue,urlcolor=black,citecolor=blue]{hyperref}

\begin{document}

\preprint{APS/123-QED}

\title{THz Second and Third Harmonic Generation in PdCoO$_2$ Thin Films}%

\author{Tim Priessnitz}%
    \email{t.priessnitz@fkf.mpg.de}%
    \affiliation{Max Planck Institute for Solid State Research, 70569 Stuttgart, Germany}%
\author{Liwen Feng}
    \affiliation{Max Planck Institute for Solid State Research, 70569 Stuttgart, Germany}%
    \affiliation{Institute of Solid State and Materials Physics, TUD Dresden University of Technology, 01069 Dresden, Germany}%
\author{Thales V. A. G. de Oliveira}%
    \affiliation{Helmholtz-Zentrum Dresden-Rossendorf, 01328 Dresden, Germany}%
\author{Graham Baker}
    \affiliation{Max Planck Institute for Chemical Physics of Solids, 01187 Dresden, Germany}
\author{Igor Ilyakov}
    \affiliation{Helmholtz-Zentrum Dresden-Rossendorf, 01328 Dresden, Germany}%
\author{Alexey Ponomaryov}
    \affiliation{Helmholtz-Zentrum Dresden-Rossendorf, 01328 Dresden, Germany}%
\author{Atiqa Arshad}
    \affiliation{Institute of Solid State and Materials Physics, TUD Dresden University of Technology, 01069 Dresden, Germany}%
    \affiliation{Helmholtz-Zentrum Dresden-Rossendorf, 01328 Dresden, Germany}%
\author{Gulloo Lal Prajapati}
    \affiliation{Helmholtz-Zentrum Dresden-Rossendorf, 01328 Dresden, Germany}%
\author{Jan-Christoph Deinert}
    \affiliation{Helmholtz-Zentrum Dresden-Rossendorf, 01328 Dresden, Germany}%
\author{Sergey Kovalev}
    \affiliation{Helmholtz-Zentrum Dresden-Rossendorf, 01328 Dresden, Germany}%
    \affiliation{Fakultät Physik, Technische Universität Dortmund, 44227 Dortmund, Germany}%
\author{Bernhard Keimer}%
    \affiliation{Max Planck Institute for Solid State Research, 70569 Stuttgart, Germany}
\author{Stefan Kaiser}%
    \email{stefan.kaiser@tu-dresden.de}
    \affiliation{Max Planck Institute for Solid State Research, 70569 Stuttgart, Germany}%
    \affiliation{Institute of Solid State and Materials Physics, TUD Dresden University of Technology, 01069 Dresden, Germany}%

\date{\today}

\begin{abstract}
    Terahertz high harmonic generation (THz HHG) is a common property of nonlinear systems. Recently it has been used to investigate fundamental principles that govern transport and nonlinear dynamics in novel quantum materials like graphene, Dirac semimetals or high-temperature superconductors. Here, we report on the observation of exceptionally large THz second harmonic and third harmonic generation in thin films of the highly conducting delafossite PdCoO$_2$ down to low temperatures. The growth of this material on offcut substrate allows for a significant enhancement of the third harmonic intensity compared to ordinary $c$-axis grown thin films. Furthermore, it appears to be a necessity for the observation of THz second harmonic generation. We model the temperature dependence of the third harmonic generation by means of Boltzmann transport theory and provide an explanation for the second harmonic generation by comparing the system to the electric field induced second harmonic generation. The present investigation thus provides an important contribution to the ongoing discussion of low temperature origins of THz HHG and might serve as a new platform for THz high harmonic applications.
\end{abstract}

\maketitle

\section{\label{sec:intro}Introduction}
    Functional materials that allow for the utilization of terahertz (THz) frequencies for modern high-speed electronics moved into the center of intense investigations over the last few years \cite{kovalev2021electrical}. A key property for the application in electronics and optoelectronics is further the possibility of high harmonic generation in this particular frequency regime. Previous findings revealed efficient THz high harmonic generation in graphene and other Dirac and Weyl semimetals, and attributed this effect to the unique electronic band structure and properties of Dirac fermions \cite{hafez2018extremely,kovalev2020non,cheng2020efficient}. The search quickly expanded to more elaborate structures involving metasurfaces on graphene and topological insulators in order to enhance the high harmonic generation \cite{giorgianni2016strong,tielrooij2022milliwatt}. In addition, many other materials have been found to exhibit strong non-linearities in the THz regime, in particular also showing THz third harmonic generation, like doped semiconductors, elemental transition metals, conventional and unconventional superconductors or correlated materials, that could potentially be utilized for THz high harmonic applications \cite{meng2023higher,salikhov2023spinorbit,matsunaga2014light,chu2020phase,kozina2019terahertz,feng2023dynamical,reinhoffer2024strong}.

    The metallic delafossite PdCoO$_2$ is one of the best conducting transition metal oxides. With a resistivity of $\rho_{ab} = 2.6\;\mathrm{\mu\Omega cm}$ at room temperature and an exceptionally large electron mean-free path, it even has a higher conductivity per electronic charge carrier than most conducting elements \cite{takatsu2007roles,hicks2012quantum,daou2017unconventional,mackenzie2017properties}. While the origin of these exceptional properties has been subject to intensive experimental and theoretical investigations, recent reports of hydrodynamic and directional ballistic transport reignited the research around this compound and its extraordinary electronic properties \cite{moll2016evidence,bachmann2022directional,lechermann2021basic,yao2024origin,baker2024nonlocal}. The confinement of the main electronic transport to two dimensions, a very high Fermi velocity, exceptionally weak electron-phonon coupling and weak electron-electron and electron-impurity scattering, as well as a high level of crystallinity were identified experimentally and theoretically to be responsible for the large electronic conductivity in PdCoO$_2$ \cite{noh2009anisotropic,hicks2012quantum,mackenzie2017properties,sunko2020controlled,lechermann2021basic,yao2024origin,baker2024nonlocal,zhang2024crystal}. A recent study further revealed that PdCoO$_2$ serves as a promising THz emitter upon optical pumping, pointing towards its application potential for ultrafast electronics and optoelectronics \cite{yordanov2023generation}.

    Here, we report on the time resolved detection of THz high harmonic generation, with a focus on the second and third harmonic, in thin films of the highly conducting delafossite PdCoO$_2$, grown on regular and offcut substrates, down to low temperatures. We further show that the utilization of offcut-grown thin films allows for the observation of THz second harmonic generation and can also enhance the third harmonic emission. To describe the temperature dependence of the third harmonic, we employ semi-classical calculations based on Boltzmann transport theory. Our results indicate that the exceptionally low scattering rate in PdCoO$_2$ is beneficial for efficient THz third harmonic generation. 

    \begin{figure}
        \includegraphics[width=\linewidth]{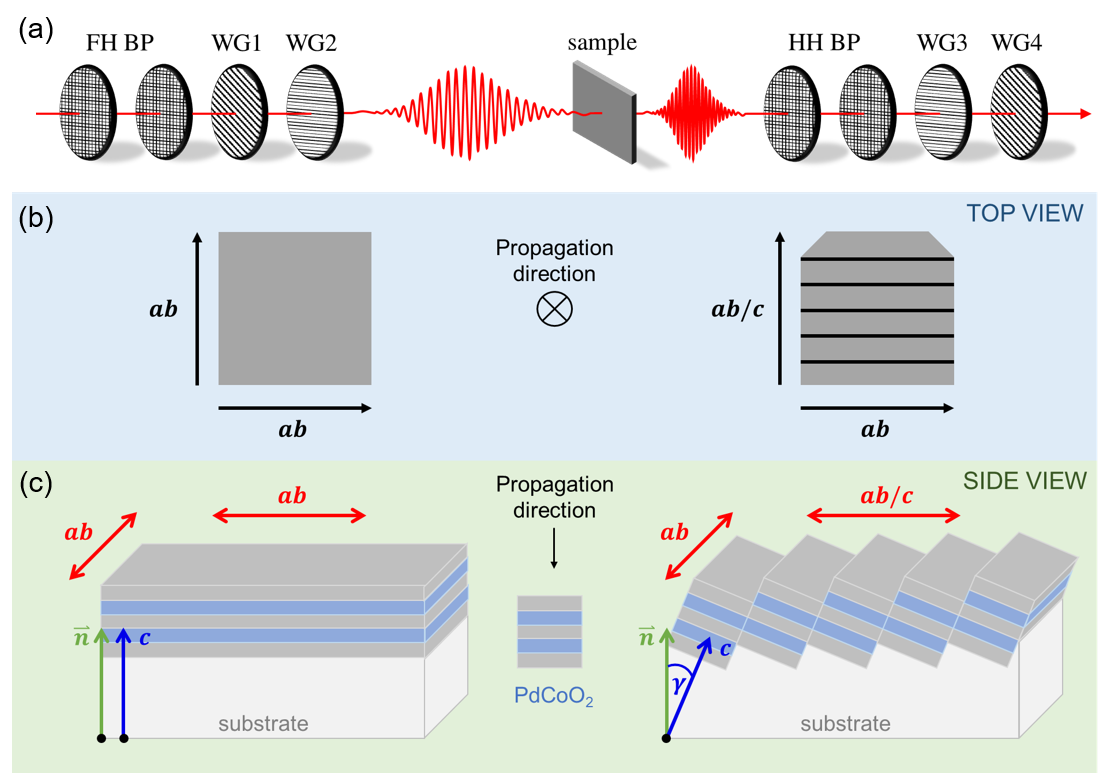}
        \caption{\label{fig:figure01setup} THz high harmonic generation from a PdCoO$_2$ thin film grown on $\gamma=10^\circ$ offcut sapphire substrate. (a) Schematics of the experimental setup with fundamental bandpass filters (FH BP) to ensure a narrow band THz pulse, wire grid polarizers (WG1-4) and high harmonic bandpass filters (HH BP). (b) Schematic top view of the thin film sample grown on regular and offcut substrate (c) Corresponding side view of the thin film samples.}
    \end{figure}
    
    \section{\label{sec:exp}Experimental Details}
    High quality PdCoO$_2$ thin films with 10 nm thickness were grown by pulsed laser deposition on $c$-axis-oriented $\gamma = 0^\circ$ (0001) and off-cut $\gamma = 10^\circ, 15^\circ$ ($0001\rightarrow1120$) Al$_2$O$_3$ substrates. Details are described elsewhere \cite{yordanov2019large,yordanov2023generation}. Top and side view of the investigated thin films are shown schematically in Figure \ref{fig:figure01setup}.
    
    The THz high harmonic experiments were carried out at the high-field high-repetition rate terahertz facility at the ELBE accelerator (TELBE) at the Helmholtz-Zentrum Dresden-Rossendorf (HZDR), which provided the required high intensity, carrier envelope phase stable, tunable and narrowband THz radiation \cite{green2016high}. A sketch of the experimental setup used is illustrated in Figure \ref{fig:figure01setup}a. The fundamental frequency (FH) is first cleaned from any other frequency components coming from the THz source by two bandpass filters (FH BP). After that, wire grid polarizers are used to control the fluence (WG1), the incident polarization on the sample (WG2), the analyzed polarization channel (WG3) and to fix the THz polarization for the electro-optic sampling (WG4). In order to increase the signal to noise ratio, additional bandpass filters for the high harmonic of interest (HH BP), i.e. second and third harmonic, are placed after the sample. Polarization dependent measurements were conducted at room temperature with the sample mounted in a rotational mount unless stated otherwise. The full $360^\circ$ polarization dependence in Figure \ref{fig:figure02} was extrapolated based on symmetry arguments. All THz signals are detected via electro-optic sampling (EOS) in a nonlinear crystal (ZnTe (110), $d=1$ mm) using a fs-laser amplifier system (Coherent RegA with pulse length $\tau_\mathrm{lp}=45$ fs), that is synchronized to the master clock of the TELBE superradiant undulator source. The fundamental frequency provided by the undulator and used in all experiments is 0.3 THz. The maximum average power of the incident THz field is $P_\mathrm{THz,avg}=28.6$ mW with a spotsize (full width half maximum) of $l=870$ $\mathrm{\mu}$m at the sample position and a repetition rate of $f_\mathrm{rep}=50$ kHz. The estimated peak electric field of the fundamental incident on the sample is $23.6$ kV/cm.
    
\section{\label{sec:results}Experimental Results}
\begin{figure}[b]
        \includegraphics[width=\linewidth]{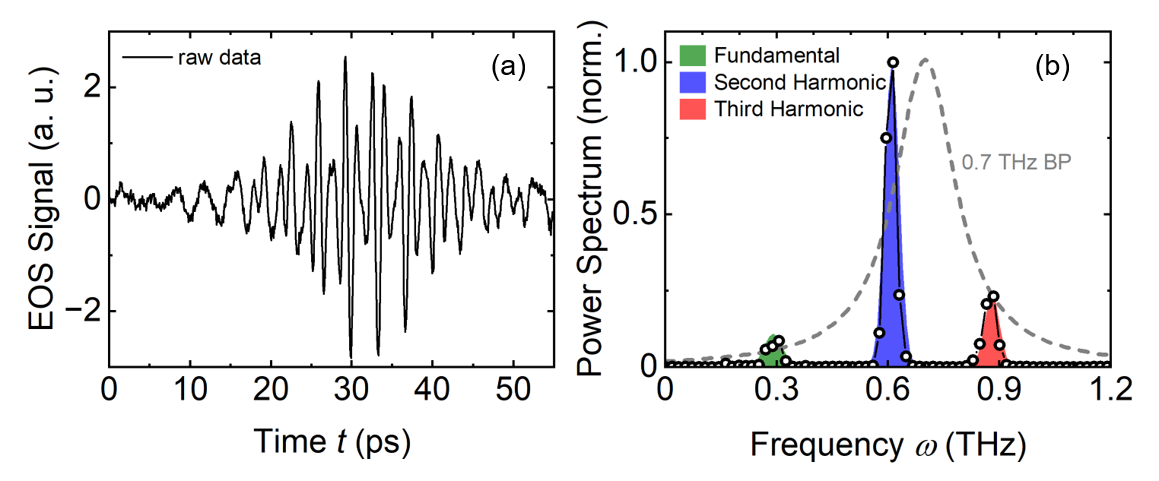}
        \caption{\label{fig:figure01} THz high harmonic generation from a PdCoO$_2$ thin film grown on $\gamma=10^\circ$ offcut sapphire substrate. (a) Signal from the electro-optical sampling (EOS) showing the THz pulse containing higher harmonic components in the time domain using two physical bandpass filters centered around 0.7 THz. (b) Corresponding power spectrum with the fundamental (green), second (blue) and third (red) harmonic highlighted. The dashed line indicates the estimated transmission profile of the used bandpass filter.}
    \end{figure}

    We performed THz high harmonic generation experiments on PdCoO$_2$ thin films grown on different offcut sapphire substrates. In Figure \ref{fig:figure01} the observed time domain signal and the corresponding power spectrum of the THz high harmonics generated in a $\gamma=10^\circ$ PdCoO$_2$ thin film at room temperature are shown. The incident polarization is parallel to the $ab/c$ direction of the sample. A 0.7 THz bandpass filter was used to make the second and third harmonic visible at the same time. Notably, if the bandpass filter just cut out a signal from the noise floor, the large second harmonic peak would be centered at the central frequency of the filter. Due to the transmission properties of the bandpass filter employed in our experiments, the detected intensity of the second harmonic significantly exceeds the one of the third harmonic. However, the intrinsic intensity of the observed third harmonic is generally at least an order of magnitude larger than the second harmonic (see Figure \ref{fig:figure04}). We further confirmed the intrinsic origin of the third harmonic by alternating the position of the physical bandpass filters in the setup (Appendix \ref{appendix:proof}). A detailed fluence dependence of the fundamental, second and third harmonic is provided in Appendix \ref{appendix:fluence}.

    \subsection{Room Temperature Polarization Dependence of the Fundamental and Third Harmonic}
\begin{figure}
        \includegraphics[width=\linewidth]{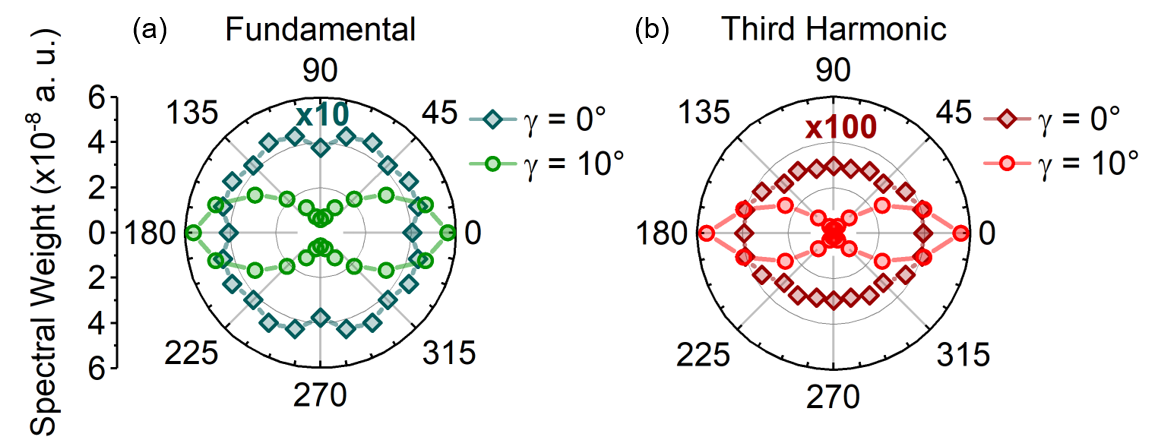}
        \caption{\label{fig:figure02} Detailed polarization dependence of the (a) fundamental transmission as well as the generated (b) third harmonic from PdCoO2 thin films grown on $\gamma=0^\circ,10^\circ$ offcut sapphire substrates at room temperature. Data from the $\gamma=0^\circ$ thin film have been enhanced for visibility as described in the legend.}
    \end{figure}

    In order to characterize the emission properties of the PdCoO$_2$ films grown on regular $\gamma=0^\circ$ and offcut $\gamma=10^\circ$ sapphire substrates, we performed polarization dependent measurements. The results are depicted in Figure \ref{fig:figure02}. We define the spectral weight for the fundamental and higher harmonics by a windowed ($\omega\pm0.1$ THz) integration around the corresponding peak in the power spectrum and use this as our observable. The angles in the polar plots are defined such that $0^\circ$ corresponds to the incident polarization being parallel to the $ab/c$ axis in the case of the film on offcut substrate. Data from the $\gamma=0^\circ$ sample are enhanced for visibility. We observe an almost isotropic response in the fundamental as well as the third harmonic for the regular, i.e. $\gamma=0^\circ$, PdCoO$_2$ thin film. In contrast, the $\gamma=10^\circ$ film shows a highly anisotropic behavior with minima in the emission for the incident polarization being parallel to the $ab$ axis of the sample and maxima for the incident polarization parallel to the $ab/c$ axis.
 
    \begin{figure*}
        \includegraphics[width=\linewidth]{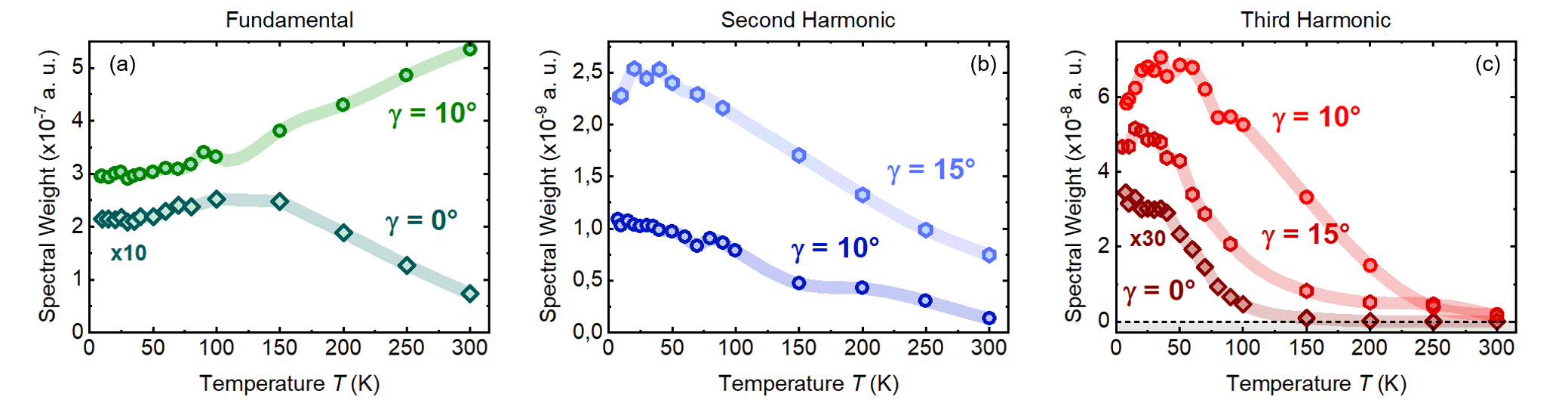}
        \caption{\label{fig:figure03} Temperature dependence of the spectral weight of the (a) fundamental transmission as well as the generated (b) second and (c) third harmonic from PdCoO2 thin films grown on $\gamma=0^\circ,10^\circ,15^\circ$ offcut sapphire substrates. Data from the $\gamma=0^\circ$ thin film have been enhanced for visibility. Notably, there is no second harmonic generation observed for the $\gamma=0^\circ$ sample.}
    \end{figure*}

    \subsection{Temperature Dependence of the High Harmonics}
    We extend our study to temperature dependent THz high harmonic generation. The main results for the observed fundamental, second and third harmonic are shown in Figure \ref{fig:figure03} for thin films of PdCoO$_2$ on $\gamma=0^\circ,10^\circ$ and $15^\circ$ offcut substrates. Samples were oriented inside the cryostat such that the incident light polarization was parallel to the $ab/c$ axis in the case of the tilted PdCoO$_2$ films. For the regular $\gamma=0^\circ$ sample, we observe a slight increase of the transmitted fundamental (FH) with decreasing temperature before it reaches a plateau below 50 K. The spectral weight of the third harmonic (TH) on the other hand remains very low upon cooling to an onset temperature around 150 K. Below this temperature the TH signal increases significantly until it also reaches a plateau below 50 K. It is worth noting that the FH signal acquired during the same temperature run as a higher harmonic is strongly distorted due to the suppression of the high harmonic bandpass filter. The temperature dependent data for the fundamental frequency behavior were therefore acquired in a separate measurement without the high harmonic bandpass filters.

    In contrast to the $\gamma=0^\circ$ response, the $\gamma=10^\circ$ PdCoO$_2$ sample shows a steady decrease of the FH with decreasing temperature until it also reaches a more or less constant value below 100 K. The TH signal for the $\gamma=10^\circ$ and $15^\circ$ increases steadily with decreasing temperature but with a broader slope than in the $\gamma=0^\circ$ case. At around 50 K a peak is reached before the spectral weight slightly decreases again. For the larger offcut angle the peak shifts to a lower temperature, at around 25 K. Additionally, while it is absent for the regular film, the offcut grown films exhibit a second harmonic (SH), which shows a gradual increase with decreasing temperature but without reaching a peak or plateau.

    \subsection{Low Temperature Polarization Dependence}
    \begin{figure}[b]
        \includegraphics[width=\linewidth]{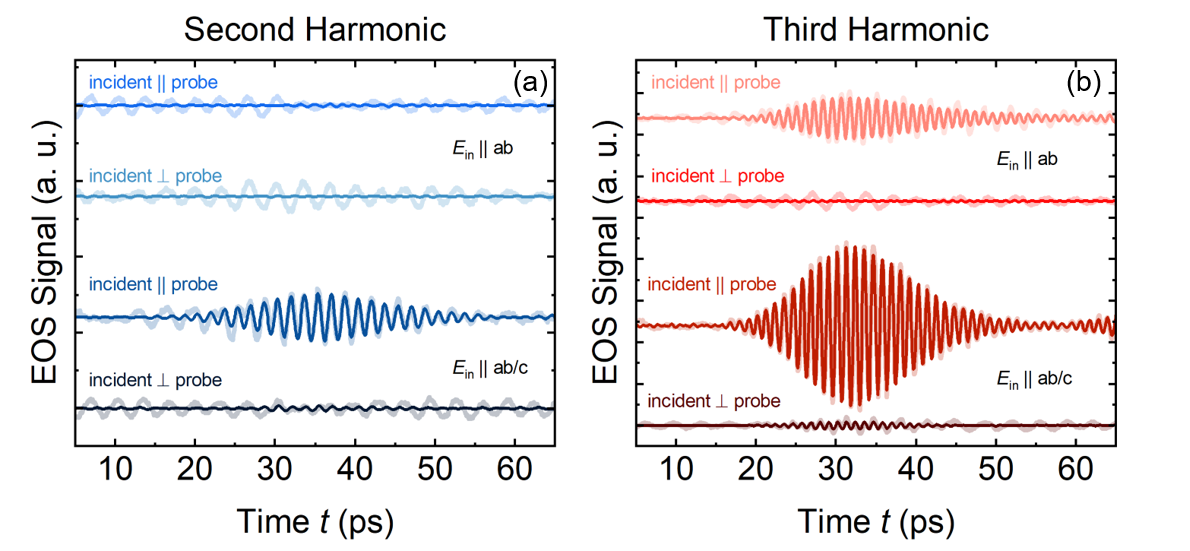}
        \caption{\label{fig:figure04} Coarse polarization dependence of the (a) second harmonic and (b) third harmonic from a PdCoO2 thin film grown on $\gamma\neq 0^\circ$ offcut sapphire substrates at $T=9$ K. The upper half of each panel represents the response for the incident polarization aligned along the $ab$-axis of the sample. The lower half of each panel shows the response for the incident polarization aligned along the $ab/c$-axis of the sample. Parallel and cross-polarized analyzing channels are indicated as “incident $\parallel$ probe” and “incident $\perp$ probe”, respectively.}
    \end{figure}
    We further investigated the polarization dependence of the second and third harmonic from a $\gamma=10^\circ$ PdCoO$_2$ film at low temperature ($T=9$ K). The experimental results are illustrated in Figure \ref{fig:figure04}. Using wire grid polarizers, we checked two different orientations of the incident THz beam, with the polarization parallel to the $ab$ plane and parallel to the $ab/c$ plane, and analyzed the parallel and perpendicular polarization channels with respect to the incident polarization. This is achieved by rotating the polarizer after the sample (WG3) such that only the parallel or perpendicular component can pass and with the detection being optimized and fixed to $45^\circ$ with the last polarizer (WG4) (see Figure \ref{fig:figure01setup}). This ensures that both parallel and perpendicular polarization are detected in a balanced fashion, meaning both polarization have the same intensity on the detector. Similarly, the incident polarization on the sample is selected by such a configuration of the polarizers in front of the sample (WG1 and WG2), keeping the incident power on the sample the same for both polarizations. The SH signal only appears when the $ab/c$ axis is probed and is absent in the cross-polarization channel, indicating that it follows the fundamental polarization. For the incident polarization parallel to the $ab$ axis of the PdCoO$_2$ sample, SH can be observed in neither of the polarization channels. For the TH, a signal can be observed for both, the incident THz polarized parallel to the $ab$ direction and parallel to the $ab/c$ direction of the sample, respectively, while it is also absent in the cross-polarized channels. However, the addition of the $c$ axis contribution significantly enhances the TH generation.

\section{\label{sec:discussion}Discussion}
    \subsection{Room Temperature Polarization Dependence}
    The polarization dependence of the $\gamma=0^\circ$ PdCoO$_2$ film as displayed in Figure \ref{fig:figure02} is as one would expect for a centro-symmetric material with space group $R\bar{3}m$ looking solely at the isotropic in-plane response. For the $\gamma=10^\circ$ sample, however, the response is clearly anisotropic. While the symmetry reduction is generally expected due to the tilt of the $c$ axis away from the surface normal, we note that from a top view, the terrace structure of such tilted thin films appears like many equally spaced metallic stripes, which is very similar to a typical wire-grid polarizer as it is commonly used for low THz frequencies. Indeed, the observed polarization dependence looks strikingly similar to the transmission profile of a wire-grid polarizer. We can thus attribute the anisotropic polarization dependence to the macroscopic texture of the sample. We further emphasize that despite their relative weakness, the fundamental and third harmonic signals $90^\circ$ away from the maximum (i.e., as the incident polarization is parallel to the $ab/c$ direction of the thin film) are comparable to those of the $\gamma=0^\circ$ sample as expected because this polarization geometry effectively probes the in-plane response.

    \subsection{Temperature Dependence of the High Harmonics}
    The temperature dependence of the fundamental for $\gamma=10^\circ$ is in agreement with the increasing conductivity upon cooling and can be thus simply attributed to a higher reflectivity at lower temperatures \cite{yordanov2019large,barbalas2022disorder}. For high temperatures the $\gamma=0^\circ$ sample shows an opposite trend, however, we note that the measured values are a factor of ten smaller than for the $\gamma=10^\circ$ sample, thus the absolute changes remain small.
    
    In order to describe the observed temperature dependence of the third harmonic as shown in Figure \ref{fig:figure03}, we follow a semi-classical approach to  calculate the nonlinear electromagnetic response of PdCoO$_2$. Solving the Boltzmann equation under the assumption of local electrodynamics and using the relaxation time approximation, we obtain the $n$th-order conductivity at the $k$th harmonic frequency for a conductor with arbitrary electronic dispersion $\mathcal{E}_k$
    \begin{eqnarray}
        \sigma^{(n)}(k\omega,T) = D^{(n)}(T)\left( \begin{array}{c} n \\ (n-k)/2)\end{array}\right)\nonumber\\
        \times\frac{1}{(1/\tau(T)-i\omega)^{(n+k)/2} (1/\tau(T)+i\omega)^{(n-k)/2}},
        \label{eq:sigma_general}
    \end{eqnarray}
    where we have defined
    \begin{eqnarray}
        D^{(n)}(T) \equiv \int d\mathcal{E}\left(-\frac{\partial f_0}{\partial\mathcal{E}}\right)M^{(n)}(\mathcal{E})
    \end{eqnarray}
    and
    \begin{eqnarray}
        M^{(n)}(\mathcal{E}) \equiv \frac{(-1)^{n-1}e^{n+1}}{n!\hbar^{n+1}} \frac{2}{(2\pi)^d}
        \int_{\mathcal{S}(\mathcal{E})} dS \frac{v_{\mathbf{k}x}}{v_\mathbf{k}} \frac{\partial^n \mathcal{E}_\mathbf{k}}{\partial k_x^n}
    \end{eqnarray}
    with $f_0$ being the Fermi-Dirac function, $\mathcal{E}_\mathbf{k}$ the electronic dispersion, $\mathcal{S}(\mathcal{E})$ a surface of constant energy, and $v_\mathbf{k}$ the velocity given by $v_\mathbf{k} = (1/\hbar)\partial\mathcal{E}_\mathbf{k}/\partial k$. Furthermore, we define the limit of $D^{(n)}(T)$ for $T\ll T_\mathrm{F}$, applicable to high-density metals like PdCoO$_2$, which is a temperature-independent value:
    \begin{eqnarray}
        D_0^{(n)} \equiv \lim_{T/T_\mathrm{F}\to 0} D^{(n)}(T) = M^{(n)}(\mathcal{E}_\mathrm{F}).
    \end{eqnarray}
    In particular, application of Equation \ref{eq:sigma_general} gives the first-order conductivity
    \begin{subequations}
    \begin{eqnarray}
        \sigma^{(1)}(\omega) = \frac{D^{(1)}}{1/\tau-i\omega},
    \end{eqnarray}
    the third-order conductivity at the fundamental frequency
    \begin{eqnarray}
        \sigma^{(3)}(\omega) = \frac{3D^{(3)}}{(1/\tau-i\omega)^2(1/\tau+i\omega)},
    \end{eqnarray}
    and the third-order conductivity at the third-harmonic
    \begin{eqnarray}
        \sigma^{(3)}(3\omega) = \frac{D^{(3)}}{(1/\tau-i\omega)^3}.
    \end{eqnarray}
    \end{subequations}

    We now consider a film of thickness $d$ on a substrate with refractive index $n$. Assuming that $d$ is much shorter than the relevant wavelength(s), the transmitted field $E_\mathrm{t}(t)$ is given by \cite{lee2020electrically,navaeipour2022effects}
    \begin{eqnarray}
       E_\mathrm{t}(t) = \frac{1}{1+n} \left[ 2E_\mathrm{i}(t) - dZ_0 J(t)\right],
       \label{eq:E_trans}
    \end{eqnarray}
    where $E_\mathrm{i}(t)$ is the incident field, $Z_0$ the vacuum impedance and $J(t)$ the total current density. Applying Equation \ref{eq:E_trans} and keeping only terms up to third order, we find that the transmitted field at the fundamental is given by
    \begin{eqnarray}
        E_\mathrm{t}(\omega) = \frac{2}{1+n+dZ_0 \left[ \sigma^{(1)}(\omega) + \sigma^{(3)}(\omega)E_\mathrm{t}(\omega)^2\right]} E_\mathrm{i}(\omega)
    \end{eqnarray}
    and the transmitted field at the third harmonic is given by
    \begin{eqnarray}
        E_\mathrm{t}(3\omega) = \frac{-dZ_0\sigma^{(3)}(3\omega)} {1+n+dZ_0 \sigma^{(1)}(\omega)} E_\mathrm{t}(\omega)^3.
    \end{eqnarray}
    We can then define the effective susceptibility
    \begin{eqnarray}
        \chi_\mathrm{eff,t}^{(3)} \equiv \left|\frac{E_\mathrm{t}(3\omega)}{E_\mathrm{t}(\omega)^3}\right|.
    \end{eqnarray}

    To finally calculate the effective susceptibility, we enter the material parameters $D_0^{(1)}$, $D_0^{(3)}$, and $\tau(T)$. $D_0^{(1)}$ and $\tau(T)$ can be inferred from existing experiments. There are two sources of temperature dependence of the nonlinear conductivity, namely $D^{(n)}(T)$ and $\tau(T)$. The work by Sun \textit{et al.} \cite{sun2018third} on third harmonic generation in graphene focused on the temperature dependence coming from $D^{(n)}(T)$. However, for PdCoO$_2$ $T_\mathrm{F}\approx3\times10^4\;\mathrm{K}$, so that for $T<300\;\mathrm{K}$, $T\ll T_\mathrm{F}$ and $D^{(n)}(T)\approx D_0^{(0)}$. Therefore, the above results would predict that the main source of the temperature dependence is the scattering time $\tau(T)$.

    To estimate the temperature dependence, we use the phenomenological temperature dependence
    \begin{eqnarray}
        \Gamma(T)\equiv\frac{1}{\tau} = \Gamma_0 + \frac{\Gamma_1}{e^{T_0/T}-1}
    \end{eqnarray}
    which captures both, the low temperature exponentially-activated behavior and the high temperature linear behavior as observed in the resistivity of bulk and thin films. From linear THz conductivity measurements on PdCoO$_2$ thin films by Barbalas \textit{et al.} \cite{barbalas2022disorder}, we estimate $\Gamma_0=1.5$ THz, $\Gamma_1=3.5$ THz, and $T_0=163$ K for our $\gamma=0^\circ$ thin film. We then use this temperature dependence to estimate the temperature dependence of $\sigma^{(1)}(\omega)$ and $\sigma^{(3)}(3\omega)$.
    
    $D_0^{(3)}$ can be calculated from a band structure model (see Appendix \ref{appendix:magnitude}), or treated as a fitting parameter which sets the absolute magnitude of $\chi_\mathrm{eff,t/i}^{(3)}$.
    
\begin{figure}
        \includegraphics[width=\linewidth]{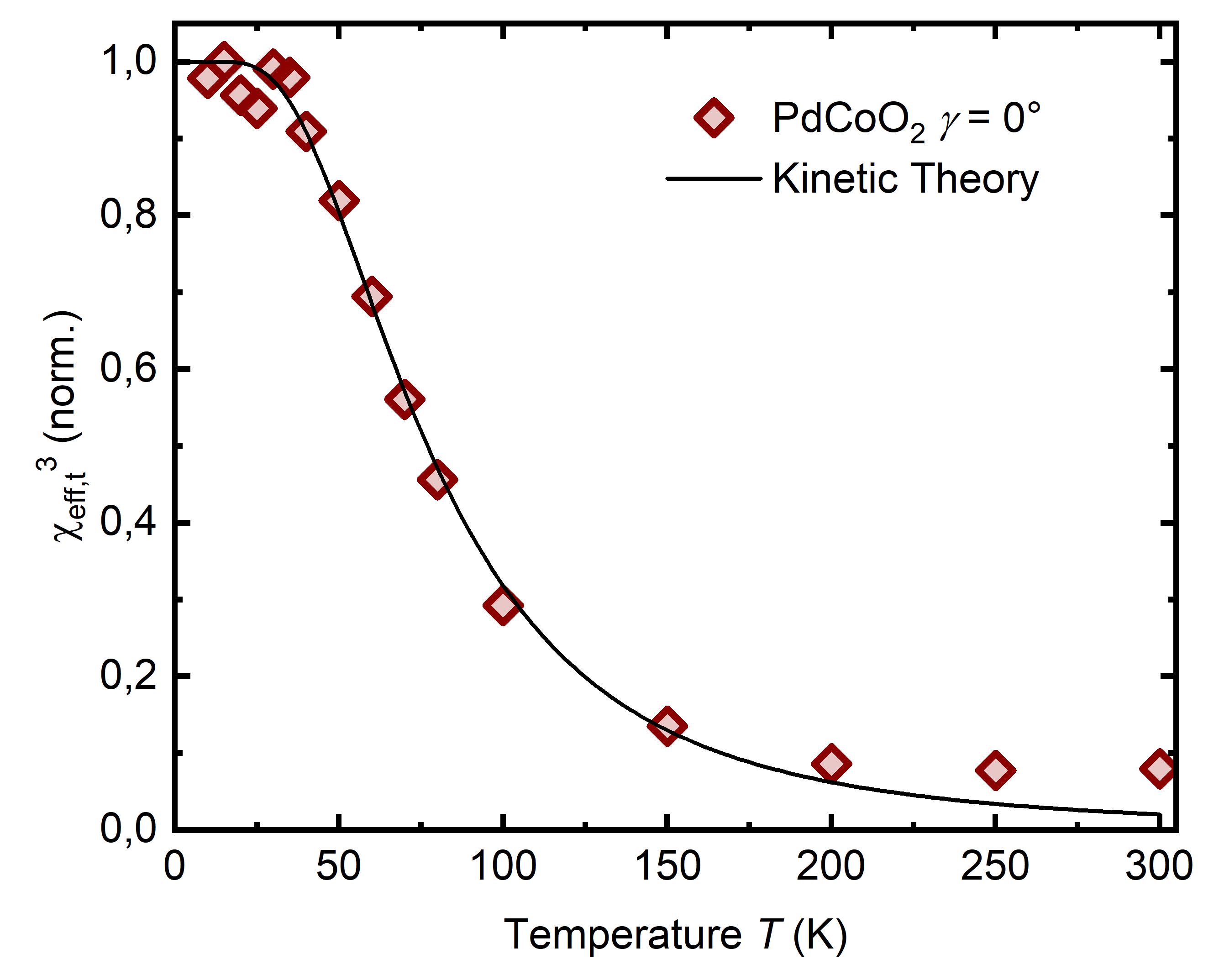}
        \caption{\label{fig:figure05} Calculated third order susceptibility as a function of temperature for a $\gamma=0^\circ$ PdCoO$_2$ film, using $\omega/2\pi=0.3$ THz, $d=10$ nm, and $n_{ab}=330$.}
    \end{figure}

    The calculated temperature dependence of $\chi_\mathrm{eff,t}^{(3)}$ is shown in Figure \ref{fig:figure05} and is in excellent agreement with the experimental data acquired from the $\gamma=0^\circ$ PdCoO$_2$ thin film.
    
    It is worth noting that the peak-like feature at low temperatures in the third harmonic signal of the tilted ($\gamma\neq0^\circ$) thin films cannot be simply explained by the remarkable metallic properties of PdCoO$_2$. While this behavior with temperature shows some similarities with predictions for exotic transport regimes such as hydrodynamics, which have also been discussed for PdCoO$_2$, their contribution in our case is unlikely since they arise from the temperature dependence of $D^n(T)$ which is temperature independent in our case where $T\ll T_\mathrm{F}$ \cite{sun2018universal,sun2018third,moll2016evidence}. Instead, this feature is likely related to the structure of the offcut film as it shifts towards lower temperatures with increasing offcut angle $\gamma$.
    
    While the general temperature behavior of the third harmonic generation in a regular, i.e. $\gamma=0^\circ$, PdCoO$_2$ thin film can be well described by kinetic theory and it basically follows the expected temperature dependence of the scattering rate in a metal, the temperature dependence of the second harmonic is nearly linear and does not just follow the scattering rate. The existence of a visible second harmonic is somewhat surprising as PdCoO$_2$ is a centrosymmetric material which usually does not exhibit any second harmonic generation consistent with the absence of any second harmonic from the $\gamma=0^\circ$ film. Together with the polarization dependence shown in Figure \ref{fig:figure04}, it becomes clear that the tilted $c$ axis structure is necessary in order to generate a second harmonic. The increase in metallicity and thus the increase in mobility of the electronic charge carriers can explain the monotonic increase of the second harmonic generation with decreasing temperature. However, the absence of a clear plateau at low temperatures sheds some doubt on the direct connection of the second harmonic generation (SHG) to the carrier mobility. This leads to the conclusion that one or more other effects must be involved in the observed SHG.
    
    There are a few possible scenarios that can lead to THz SHG, which have been previously studied in materials and devices based on semiconductors or graphene, for example electric field induced SHG (or EFISHG), lattice resonances, asymmetrically designed quantum wells or damped Bloch oscillations which were frequency modulated by an externally applied field \cite{mayer1986far,bewley1993far,heyman1994resonant,winnerl2000frequency,terhune1962optical,cai2011electrically,kang2014electrifying,timurdogan2017electric,lee2020electrically}. Another aspect to consider is that the THz fields are presumably enhanced due to the intrinsic material structure, i.e. alternating stacks of highly conducting (Pd) and insulating (CoO$_6$) layers, which might act similar to the well understood field enhancement structures that effectively are comprised of nanometer-sized antennas deposited on semiconducting or dielectric substrates. These structures are known to lead to significant field enhancement factors in the range of several orders of magnitude for low THz frequencies and antennas with a few nanometers in size \cite{seo2009terahertz,park2010terahertz,park2011terahertz,merbold2011second,novitsky2012non,suwal2017nonresonant}. This would further explain why the observed third harmonic generation (THG) is much stronger in the case of the offcut films $\gamma\neq0^\circ$ compared to the THG from the regular $\gamma=0^\circ$ film.

    In the present case, the tilted structure of the naturally layered system breaks inversion symmetry. The structural constraint on the charge carrier drift in response to an external electric field is highly dependent on the direction resulting in an effective second-order susceptibility $\chi^{(2)}_\mathrm{eff}$. This also facilitates in the fact that the polarity of the second harmonic signal changes its sign upon a $180^\circ$ in-plane rotation of the tilted thin film as shown in Figure \ref{fig:appendix_fig02}. Notably, this behavior was also observed for the the directionally dependent THz generation based on a transverse thermoelectric effect in offcut-grown PdCoO$_2$ thin films \cite{yordanov2023generation}.
    
    Similar behavior is found in static field induced SHG, where an externally applied DC voltage leads to a broken symmetry in the current of photo-excited carriers in a semiconductor \cite{lee2020electrically}. In such a case the polarity of the second harmonic signal also changes depending on the sign of the applied bias $E_\mathrm{bias}$ as the effective second-order nonlinear polarization is given by $P^{(2)}_\mathrm{eff}\approx E_\mathrm{bias}\chi^{(3)}(2\omega;\omega,\omega,0)E_\mathrm{i}^2$. Furthermore, since the material inherently also exhibits third harmonic generation, it was observed that the amplitude of the third harmonic decreases while the one of the second harmonic increases with increasing bias \cite{winnerl2000frequency,lee2020electrically}. Analogous, the polarization for the present case can be approximated by $P^{(2)}_\mathrm{eff}\sim \sin(\gamma)\chi^{(3)}(2\omega;\omega,\omega,0)E_\mathrm{i}^2$, where $\sin(\gamma)$ takes over the role of $E_\mathrm{bias}$ in the case of the semiconductor device. Moreover, it can be seen from Figure \ref{fig:figure03} that the spectral weight of the second harmonic increases while the one of the third harmonic decreases with increasing offcut angle $\gamma$, well in agreement with the observations made for EFISHG.
    It is worth mentioning that the EFISHG saturates at higher bias voltages, due to the strong analogy suggesting that the SHG might also saturate at higher offcut angles.
\vspace{-0.25cm}    
\section{\label{sec:conclusion}Conclusion}
    We observed THz second and third harmonic generation in PdCoO$_2$ thin films grown on regular and offcut substrates. THz THG is well captured within the framework of kinetic theory and is thus mainly attributed to the fact that PdCoO$_2$ is a very good metal. We further showed that the offcut grown structure can be utilized to enhance the THG and also makes SHG possible. The growth of highly conducting layered materials on offcut substrates might thus serve as a novel platform for versatile THz high harmonic applications, including their integration into heterostructures with other transition metal oxides that host collective excitation phenomena receptive to such THz fields.
\vspace{-0.25cm}    
\begin{acknowledgments}
    We would like to thank Manuel Knauft for support with the semi-classical calculations, and Michael Fechner, Gideok Kim, Min-Jae Kim and Dirk Manske for fruitful discussions. Parts of this research were carried out at ELBE at the Helmholtz-Zentrum Dresden - Rossendorf e. V., a member of the Helmholtz Association. This work was supported by the German Research Foundation (Deutsche Forschungsgemeinschaft, DFG, Project No. 107745057 - CRC/TRR 80, subproject G8).
\end{acknowledgments}

\appendix
    
\section{Additional proof for THz SHG and THG}\label{appendix:proof}
    \begin{figure}[b]
        \includegraphics[width=0.8\linewidth]{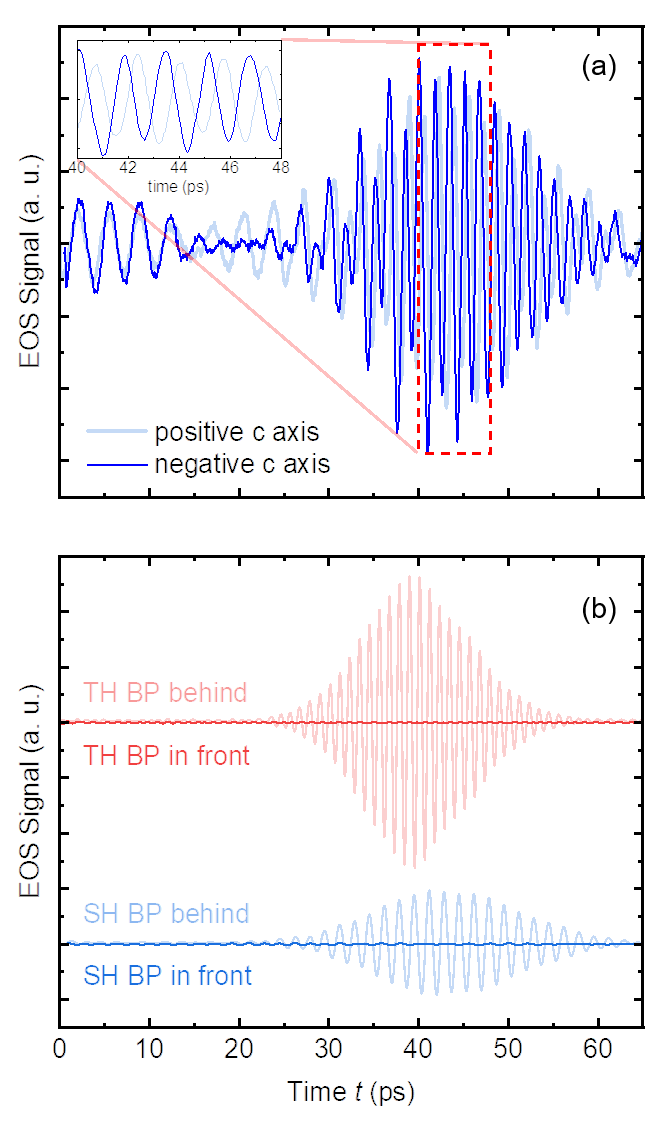}
        \caption{\label{fig:appendix_fig01} (a) Time domain EOS signals of the observed SH. Positive and negative $c$-axis indicate an in-plane rotation of the sample by $180^\circ$ which effectively results in the terraces of the tilted thin film facing the opposite direction causing a sign change of the polarity. (b) EOS signals with the SH and TH bandpass (BP) filters in front and behind the sample.}
    \end{figure}
    To provide further proof that the observed THz high harmonic signals, in particular the SH and TH, are not due to source leakage and other setup related effects, but only from the PdCoO$_2$ sample, we conducted additional measurements. First, due to the terrace structure of the offcut-grown thin films, the effective $c$-axis direction will change upon a $180^\circ$ in-plane rotation of the sample. With this, the polarity of the any emitted field from the sample in relation to this axis contribution changes its sign. This can be clearly seen in Figure \ref{fig:appendix_fig01}a and the inset. Furthermore, if the the observed harmonics are artifacts from source leakage or other setup related effects, the order of sample and high harmonic bandpass filters will not affect the observed SH and TH. However, the absence of any SH or TH signal upon placing a high harmonic bandpass in front of the sample (Figure \ref{fig:appendix_fig01}b), clearly shows that the observed signals are indeed coming from the sample.

\section{Fluence Dependence}\label{appendix:fluence}
    \begin{figure}[b]
        \includegraphics[width=\linewidth]{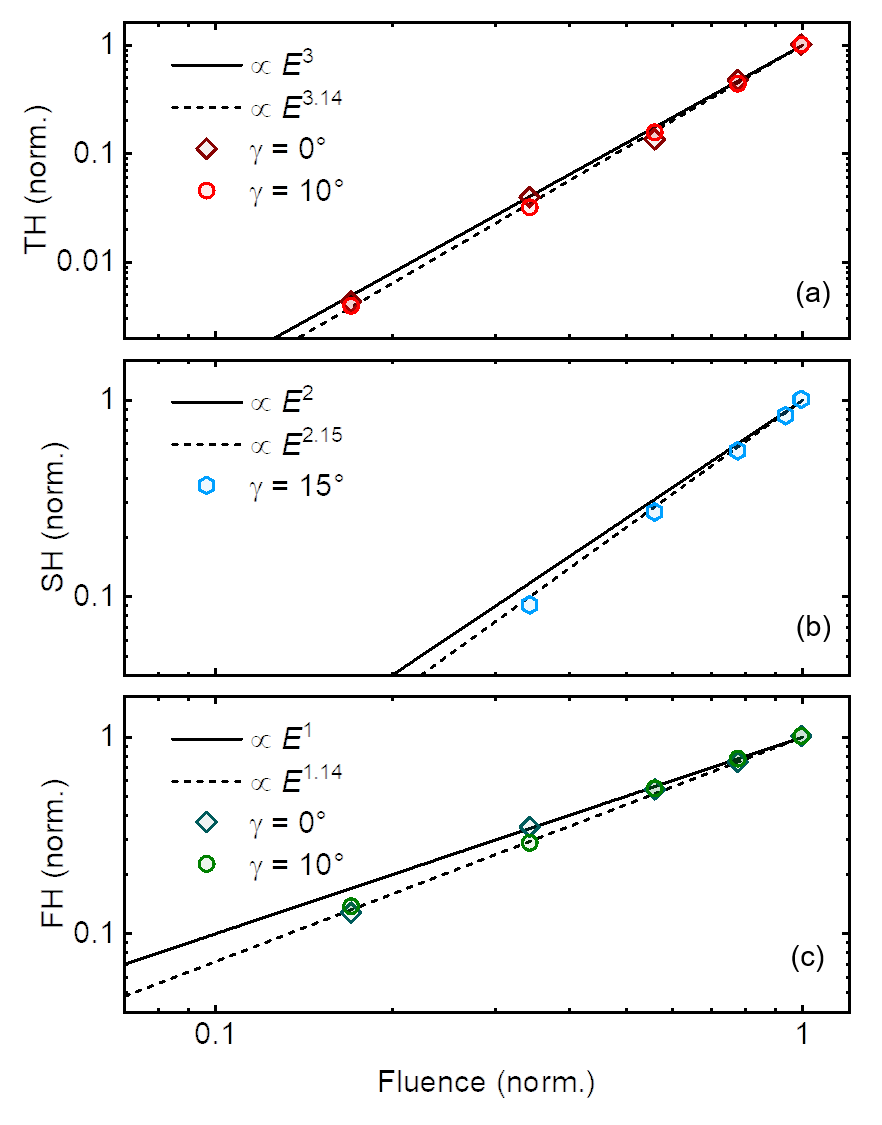}
        \caption{\label{fig:appendix_fig02} (a-c) Dependence of the normalized fundamental (FH), second harmonic (SH) and third harmonic (TH) intensities on the incident fluence with fitted power-law behavior (dashed lines). All signals slightly deviate from a pure perturbative (non)linear response (solid lines).}
    \end{figure}
    We measured the signals for the fundamental (FH), second harmonic (SH) and third harmonic (TH) for different pump fluence, i.e. for different powers of the incident THz pulse, at room temperature. The results are summarized in Figure \ref{fig:appendix_fig02}. Notably, none of signals strictly follow a linear, square or cubic law, but instead all show slight deviations and a different power-law dependence, revealing a not strictly perturbative nonlinear response.
    
    Since all deviations from the pure perturbative regime are of the same order, we speculate that the origin for this difference must come from the same process. We like to point out that such deviations from a pure perturbative fluence dependence have also been observed in other materials like graphene, Cd$_3$As$_2$ or CaRuO$_3$, where this was attributed to the dynamics of Dirac fermions and heavy quasiparticles \cite{cheng2020efficient,kovalev2020non,reinhoffer2024strong}.
    
\section{ $D_0^{(n)}$ Magnitude Estimation for {PdCoO$_2$}}\label{appendix:magnitude}
            We now turn to calculate the optical weight $D_0^{(n)}$, which sets the magnitude of the $n$th-order nonlinear conductivity. As an approximation for the electronic dispersion of PdCoO$_2$, we use a tight-binding model with nearest-neighbor and third-nearest-neighbor hopping on a triangular lattice:
            \begin{eqnarray}
                \mathcal{E}_\mathbf{k} = -2t\left[\cos(k_xa)+\cos(k_x a/2 + \sqrt{3} k_y a/2) \right.\nonumber\\
                \left. + \cos(k_x a/2 - \sqrt{3} k_y a/2) \right]\nonumber\\
                -2t^\prime\left[\cos(2k_xa)+\cos(k_x a + \sqrt{3} k_y a) \right.\nonumber\\
                \left. + \cos(k_x a - \sqrt{3} k_y a) \right]
            \end{eqnarray}
            with $t^\prime/t=0.15$. By discretizing the Fermi surface, we calculate $D_0^{(n)}$ numerically and find that
            \begin{eqnarray}
                \frac{D_0^{(3)}}{D_0^{(1)}}\approx -0.64\frac{e^2a^2}{6\hbar^2}.
            \end{eqnarray}
            Using the in-plane lattice constant of PdCoO$_2$, $a\approx2.83\;\mathrm{\AA}$, we find that
            \begin{eqnarray}
                \frac{D_0^{(3)}}{D_0^{(1)}} \approx -20\times10^9\;\mathrm{m^2/(s^2V^2)}.
            \end{eqnarray}
    
\bibliography{apssamp}

\end{document}